\documentstyle[aps,prl,twocolumn,psfig]{revtex}
\author{Thibaut Jonckheere, Beno\^\i t Gr\'emaud and Dominique Delande}
\address{Laboratoire Kastler-Brossel, Universit\'e Pierre et Marie Curie\\
Tour 12, Etage 1, 4 Place Jussieu, F-75005 Paris}
\title{Spectral properties of non-hydrogenic atoms in weak external fields}
\date{\today}

\begin{document}

\maketitle

\begin{abstract}
We study how the ionic core in a non-hydrogenic atom modifies
the dynamics of a Rydberg electron in the presence of a weak
static external field. We show
that such a system is neither regular nor chaotic: 
its energy levels display unusual
statistical properties, intermediate between the standard
Poisson and Random Matrix ones. The ionic core acts as
a scatterer whose size is comparable
to the de Broglie wavelength of the electron, inducing
specific quantum effects.
\end{abstract}

\draft
\narrowtext

The hydrogen atom in the presence of
an external magnetic field is a quantum system whose corresponding
classical dynamics is either regular or chaotic, depending on
the scaled
energy $\epsilon = E \gamma^{-2/3,}$
($E$ is the energy and $\gamma$ the magnetic field
in atomic units), and thus constitutes
a prototype for studies on quantum chaos \cite{physrep}.

In correspondance with the evolution of the classical dynamics
from quasi-integrable for a weak magnetic field
(low negative scaled energy $\leq -0.5$) to chaotic at high field
(high scaled energy $\geq -0.13$),
the statistical properties of the energy levels
evolve from a Poissonian spectrum to those of the Gaussian Orthogonal
Ensemble (GOE) of random matrices \cite{physrep}. In the first case,
the nearest-neighbor spacing (NNS)
distribution -- which measures the probability to find the next
energy level at distance $s$ (in unit of the mean level
spacing) -- is $P(s)=\exp (-s),$ and in the second case, it
is approximately given by the Wigner distribution 
$P(s) =\pi s/2 \exp (-\pi s^2/4)$.

In a non-hydrogenic atom -- for example an alkali atom as used in
several experiments \cite{held,rmat,courtney} -- the Rydberg electron is in a 
highly excited
state while all
other electrons are in low excited states.
The system (nucleus+inner electrons) can be considered as
a frozen ionic core with spherical symmetry, the size of few Bohr radii.
When it is outside this ionic core, the Rydberg electron experiences
a Coulomb field created by a charge $Z$=1 at the nucleus. It is only when
it penetrates the core that it feels a stronger force. At the scale of the
Rydberg electron (few thousands Bohr radii), the ionic core appears as a very 
small object perturbing the hydrogenic dynamics. 

A non-trivial question -- addressed in this letter -- is to determine
the spectral properties of a non-hydrogenic atom 
in an external field. 
When the classical hydrogenic dynamics is chaotic,
it seems that they remain described
by the GOE \cite{courtney}. The situation is totally different when the hydrogenic
dynamics is regular because the ionic core breaks the quasi-integrability.
Courtney et al \cite{courtney} 
have shown a dramatic change in the statistical properties of energy
levels which seem to be close to the GOE ones in the presence of the ionic core.
They attribute this phenomenon to ``core induced chaos"; we here show that
this point of view is incomplete.

An efficient numerical method based on $R$-matrix theory exists for finding
the energy levels of a non-hydrogenic atom in an external field \cite{rmat}.
It relies on the physical picture outlined above: one splits
space in a outer region where the effect of the ionic core is negligible
and where the Schr\"odinger equation for the Rydberg
electron is solved
by expansion on a suitable basis, and an inner region where the 
quantum dynamics is
dominated by the ionic core (the effect of the external field is negligible)
and can be described by a set of quantum defects. The matching between 
the solutions in the two regions gives the energy levels of the non-hydrogenic
atom.

\begin{figure}
\centerline{\psfig{figure=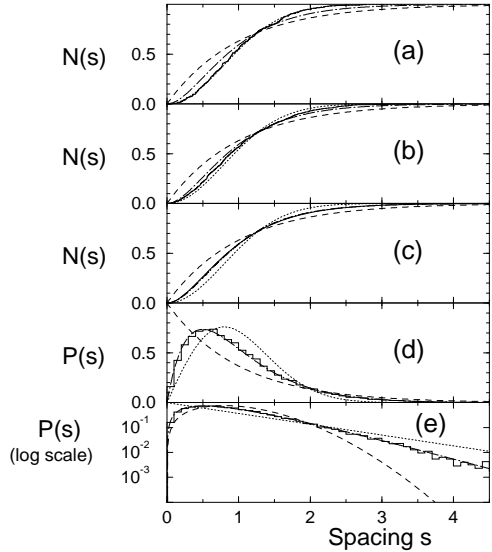,width=10cm,angle=-90}}
\medskip
\caption{Statistical properties of the energy levels of a
non-hydrogenic atom (quantum defect $\delta_{\ell=0}=0.5)$
in a weak magnetic field (scaled energy -0.5, $L_z$=0, even
parity). (a) Cumulative nearest-neighbor spacing distribution
between the $200^{th}$ and the $700^{th}$ excited states
(solid histogram), compared with the Poisson (dashed line),
semi-Poisson (dash-dotted line) and Wigner (dotted line) distributions; 
(b) and (c) Same as (a), respectively in the ranges $1000^{th}$ to $2000^{th}$
and $5000^{th}$ to $40000^{th}$ excited states; 
(d) and (e) The nearest-neighbor
spacing distribution itself, on linear and logarithmic scales (data as in c).
While, for low excited states, the distribution is well described by the Wigner
distribution, the distribution
for higher excited states is quite different, with a linear level repulsion
like Wigner and an exponential tail like Poisson. The agreement with
the intermediate semi-Poisson distribution is excellent, see especially the
tail at large spacing in (e).
In the corresponding hydrogenic situation, all the distributions are
close to the Poisson distribution.}
\label{fig1}
\end{figure}

We have calculated the lowest 40000 states of the hydrogen atom in a magnetic
field at scaled energy $\epsilon=-0.5$ (for the $L_z=0$, even parity series)
and the same set for the simplest non-hydrogenic atom, imposing
a single non-zero quantum defect $\delta_{\ell=0}=0.5$. This is
not too far from a real atom like Lithium where $\delta_0=0.4$
and $\delta_{\ell \ge 2} \ll 0.01.$
Figure~\ref{fig1} shows the cumulative spacing
distribution $N(s)=\int_0^s{P(x)\ dx}$ obtained for various
sequences of energy levels (for $\delta_{\ell=0}=0.5)$,
all obtained at the same scaled energy, that is for the same 
hydrogenic dynamics.
The lowest sequence is well described by the Wigner
distribution, but, for higher excited states, the
distribution deviates from Wigner and tends
to a well defined limit which presents level repulsion -- like 
the Wigner distribution -- at the
origin,
but also has a much longer tail for large spacings -- like the
Poisson distribution, see Figs.~\ref{fig1}(d) and (e).
Clearly, the behaviour at large $s$ is exponential, {\em not}
Gaussian.
For comparison, we have also plotted the ``semi-Poisson" distribution
(defined below) \cite{bogo}:
\begin{equation}
P(s) = 4s\ \exp (-2s),
\label{semip}
\end{equation}
which is in excellent agreement with our numerical results.
We have carefully checked that the numerically obtained distribution
does not further change for higher excited states.
In other words, when we go to higher and higher Rydberg states at fixed
scaled energy, the statistical properties of the energy levels 
{\em do not} tend to the ones of the GOE.

\begin{figure}
\centerline{\psfig{figure=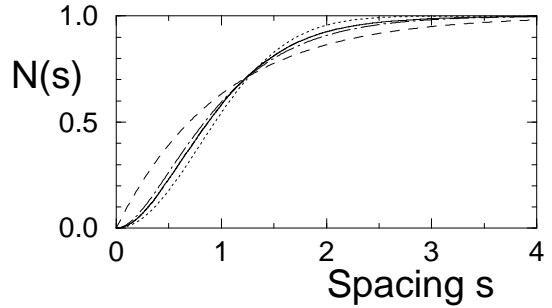,width=8cm,bbllx=110pt,bblly=150pt,bburx=430pt,bbury=340pt}}
\medskip
\caption{Same as figure 1, but in a weak electric field (scaled energy
-2.2) for states ranging from the $20000^{th}$ to the $40000^{th}$ excited
states. Again, the cumulative spacing distribution is clearly different from
the Poisson and the Wigner distributions, in good qualitative agreement with
the semi-Poisson prediction at small and large $s.$}
\label{fig2}
\end{figure}

A similar phenomenon is observed in an external 
static electric field. For scaled energy
$\epsilon=EF^{-1/2}$ (where $F$ is the electric field
in atomic units) sufficiently below the classical ionization threshold
$\epsilon=-2$, the highly excited states
are quasi-discrete. In the absence of an ionic core, the system
is integrable and displays a Poisson statistics. For a non-hydrogenic atom,
the NNS distribution is linear at small spacing (the cumulative 
distribution starts as $s^2$) and falls exponentially at large $s$ --
see Fig.~\ref{fig2}. Although it slightly deviates from
the semi-Poisson distribution (see discussion below), 
it has the same qualitative behaviour
at small and large $s$ and is clearly very different from a Wigner
distribution.
We expect such a behaviour to be general for non-hydrogenic atoms in any 
external field weak enough to keep the classical hydrogenic dynamics
quasi-integrable.

A complete understanding requires a careful analysis
of the physical phenomena taking place in the vicinity
of the ionic core. The crucial point is to know whether a semiclassical 
approximation may be used there: we argue in the following that the
answer is negative and that a specific approach is needed.

The ionic core acts as a spherical obstacle
which scatters the Rydberg electron. The size of
the scattering object (few Bohr radii) 
is comparable to the de Broglie wavelength of the Rydberg electron close
to the core, even
if one considers extremely highly excited states: the ionic core 
{\em cannot} be treated as a classical object \cite{cs}.
The effect of a spherical
scatterer is accounted for by a set of phase shifts in the different 
spherical channels. Here, these phase shifts are nothing but the quantum 
defects. Different model potentials for the ionic core may give the same 
set of quantum defects and consequently the same
spectra in presence of an external field. They will however lead to 
different classical dynamics \cite{main} and consequently 
to different thresholds for core-induced chaos: this is because near the 
nucleus, where the core strongly affects the classical trajectories and 
makes them to depend sensitively on the initial conditions, the semiclassical 
approximation breaks down. It is thus impossible to build a solution of 
the Schr\"odinger equation which follows closely one such classical 
trajectory, the wavepacket being unavoidably scattered near the
origin \cite{delos}. 
However, the breakdown of the semiclassical approximation 
is not extremely severe because the de Broglie wavelength
is comparable -- not much larger -- to the size of the ionic core.
A semiclassical approximation using 
core-scattered orbits thus give a correct qualitative 
understanding of the physics, for example the existence of additional 
modulations in the density of states and density of oscillator strengths
\cite{main}.
If quantitative results are needed  -- for example the amplitudes and phases 
of these modulations -- a quantum {\em ad hoc} treatment of the
ionic core is required, as done in \cite{cs}.
From a temporal point of view, a typical classical trajectory
follows an invariant torus of the hydrogenic motion, being from time to time
scattered by the ionic core to another torus. The
time interval between two consecutive scattering events is of the
order of the Heisenberg time, $\hbar$/(mean level spacing), which
expresses that the quantum dynamics can be considered neither as chaotic,
nor as regular: it is an intermediate situation.

The physics is very similar to the one of a rectangular two-dimensional billiard
containing a small circular scatterer, whose size is 
tuned to the de Broglie wavelength \cite{billiard}. 
While, for a relatively large obstacle, the system (a standard Sinai billiard)
has the statistical properties of the GOE, 
in the limiting case of a point scatterer, these
are somewhat different: there is linear level repulsion
at small spacing, but the tail significantly deviates from a Gaussian,
{\em exactly} as our numerical 
observations on non-hydrogenic atoms. Because of particular attention on 
the level repulsion phenomenon, the distributions obtained in \cite{billiard} 
have been incorrectly interpreted as Wigner distributions and thus as 
proofs of the existence of ``wave chaos" of ``chaos induced by quantization". 

The analogy between a non-hydrogenic atom and such a billiard 
can be made more formal. 
Suppose we are able to solve exactly the hydrogenic problem in the presence 
of an external field. $E_i$ will denote the energy levels and 
$\psi_i({\bf r})$ the eigenstates. The 
Green's function is:
\begin{equation}
G(E,{\bf r},{\bf r'})= \sum_i{\frac{\psi_i^*({\bf r'}) \psi_i({\bf r})}
{E - E_i}}.
\end{equation}
At any energy $E$, $G(E,{\bf r},{\bf r'})$ 
(viewed as a function of ${\bf r}$) 
is a solution of the Schr\"odinger equation 
for the electron, decreasing at infinity. 
Moreover, except in the immediate vicinity of the nucleus, it is also 
a solution of the Schr\"odinger equation for the non-hydrogenic atom.
Hence, the energy spectrum of the non-hydrogenic atom in the presence 
of the external field is simply given by those values of $E$ for which 
$G(E,{\bf r},{\bf r'})$ near ${\bf r}=0$ can be matched with the
field-free non-hydrogenic eigenfunctions (which depend on the quantum
defects) described in \cite{seaton}.
In the specific case of a single non-zero quantum defect 
$\delta=\delta_{\ell=0}$, 
only the $\ell=0$ spherical component plays a role as the other ones 
cancel at the origin, and we finally end with the equation:
\begin{equation}
\sum_i{\frac{|\psi_i({\bf 0})|^2}
{E_i - E}} = \frac{\cot(\pi\delta)}{2},
\label{eqseba}
\end{equation}
where the sum over $i$, which is formally divergent, has to be
conveniently renormalized. This renormalization process is somewhat
tricky and will be explained -- together with the general case of
several nonzero quantum defects -- in detail elsewhere \cite{tjdd:tbp}.
It is analogous to the one used in \cite{billiard} for a point scatterer
in a billiard (for which a similar equation is obtained for the energy
spectrum), taking into account the presence of the Coulomb potential.

Eq.~(\ref{eqseba}) not only allows 
to calculate the non-hydrogenic energy levels 
as soon as the hydrogenic ones are known, but it also gives much insight 
in the statistical properties of the energy levels. 
A first remark is that there is exactly 
one root between two poles, i.e. one non-hydrogenic energy level 
in the interval limited by two consecutive hydrogenic levels. 
We now concentrate on the case $\delta=0.5$ where the rhs in 
Eq.~(\ref{eqseba}) is zero.
The statistical properties of the roots of Eq.~(\ref{eqseba})
can be obtained from the statistical properties of the
hydrogenic energy levels $E_i$ and the numerators $|\psi_i({\bf 0})|^2$.
For a quasi-integrable situation, the $E_i$ have a
Poisson distribution, but the distribution of the numerators is not universal. 
It is then easy to show that the NNS distribution of
the roots of Eq.~(\ref{eqseba}) decreases like $\exp (-2s)$
at large distance (unlike the Wigner distribution) and
presents a linear level repulsion $P(s)=\alpha s$
at small $s$ with the coefficient $\alpha$ depending
on the distribution of the numerators. When all numerators
are equal, one gets $\alpha=\pi \sqrt{3}/2$, significantly
smaller than the semi-Poisson prediction $\alpha=4$, see Eq.~(\ref{semip})
\cite{bogo}. 
This is the situation realized for a
weak external electric field where the eigenstates are separable 
in parabolic coordinates,
and their overlaps with the usual spherical hydrogenic
states are given by a Clebsch-Gordan coefficient, which is the same
for all states in the case $\ell=m=0:$ the slope near $s=0$ is
smaller than 4, see Fig.~\ref{fig2}, and a
global small deviation from semi-Poisson is observed. For a non-uniform
distribution of the numerators, $\alpha$ increases, being for usual
smooth distributions close to 4, either smaller
or larger. This is the case for an external magnetic field 
where the numerators are widely spread and the global agreement with 
semi-Poisson almost perfect.
In order to understand qualitatively the spectral properties,
Bogomolny et al \cite{bogo} have introduced a simple model -
- baptized ``semi-Poisson model" --for
solving approximately Eq.~(\ref{eqseba}),
where each zero is lying exactly in the middle of two consecutive poles. 
The corresponding NNS distribution  -- called semi-Poisson distribution --
is given by Eq.~(\ref{semip}) and reproduces well our numerical results,
see figs.~\ref{fig1} and \ref{fig2}.

\begin{figure}
\centerline{\psfig{figure=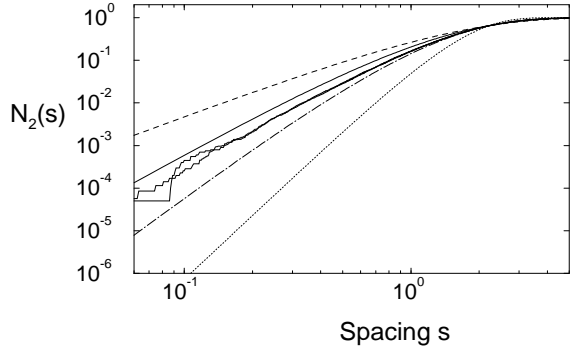,width=9cm,angle=-90}}
\medskip
\caption{Cumulative next-nearest-neighbor spacing distributions (on a
double logarithmic scale) for a
non-hydrogenic atom in a weak magnetic (data of figure 1c) or electric
(data of figure 2) field, compared with the Poisson (dashed line),
semi-Poisson (solid line), short-range Dyson model (dash-dotted line) and
GOE (dotted line) distributions. 
The functional dependance at small spacing $s$ is clearly 
proportional to $s^3,$ in agreement with the semi-Poisson model, and
deviates from all other distributions.}
\label{fig3}
\end{figure}

Several models based on different physical grounds predict linear level
repulsion. For example, the ``short-range Dyson model" discussed in
\cite{bogo} (where there is level repulsion only between adjacent states, 
in contrast with the GOE where there is level repulsion between 
any pair of states)
predicts a NNS distribution exactly equal to the semi-Poisson distribution,
although other statistical quantities behave differently from the ones
of the semi-Poisson model.
In order to discriminate between the various models, we study the
next-nearest-neighbour distribution $P_2(s)$ i.e. the statistical 
distribution of spacing between states $n$ and $n+2.$ The corresponding 
cumulative distribution $N_2(s)=\int{P_2(x)\ dx}$ is shown in Fig.~(\ref{fig3}) 
on a double logarithmic scale, in comparison with the predictions of the
Poisson, semi-Poisson, short-range Dyson and GOE models,
which behave as $s^2$, $s^3$, $s^4$ and $s^5$ respectively.
Clearly, the numerical results scale as $s^3$ in good agreement with the
semi-Poisson model, in sharp contrast with some other systems displaying 
intermediate level statistics \cite{bogo,rhombus}, which are 
closer to the short-range Dyson model. Note that semi-Poisson model
does not predict correctly the coefficient of $s^3$. Like the NNS, there is
no universality here and the model is too crude. 

When the quantum defect $\delta$ in the $\ell=0$ channel is not equal to 0.5,
the rhs in Eq.~(\ref{eqseba}) is not zero and, in average,
the roots are not
in the middle of consecutive poles. 
For equally spaced $E_i$ and equal numerators, Eq.~(\ref{eqseba}) 
can be exactly solved, and the roots
lie exactly at fraction $\delta$ of each interval.
To describe the general situation (not equally spaced $E_i$),
we define a $\delta-$Poisson model where
each level lies at $\delta E_i+(1-\delta)E_{i+1}$. The corresponding
spacing distribution is:
\begin{equation}
P_{\delta}(s)=\frac{1}{1-2\delta}\left[\exp\left(-\frac{s}{1-\delta}\right)
-\exp\left(-\frac{s}{\delta}\right)\right].
\label{deltapoisson}
\end{equation}

\begin{figure}
\centerline{\psfig{figure=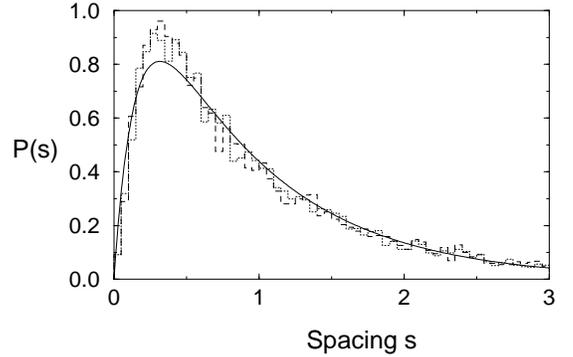,width=9cm,angle=-90}}
\medskip
\caption{Spacing distributions for a non-hydrogenic atom in a weak magnetic
field (scaled energy -0.5, from $13000^{th}$ to $22000^{th}$ excited states) 
with quantum defect $\delta_{\ell=0}=0.15$
(dotted line) and 0.85 (dashed line) in comparison with the analytical
prediction of the $\delta$-Poisson model, Eq.~(\protect{\ref{deltapoisson}}).
The excellent agreement validates the model which smoothly evolves
from a Poisson distribution for $\delta=0$ to a semi-Poisson
distribution at $\delta=0.5$ where the effect of the ionic core is maximum.}
\label{fig4}
\end{figure}

It is shown in Fig.~(\ref{fig4}) in comparison with the
NNS distributions obtained numerically for a non-hydrogenic atom in a
magnetic field with $\delta=0.15$
and 0.85. The agreement is excellent.
The same distribution is obtained for quantum
defects $\delta$ and $1-\delta,$ as predicted by Eq.~(\ref{deltapoisson}).

In conclusion, we have shown that, for a non-hydrogenic atom in a weak
external field, the presence of the ionic core which scatters the
Rydberg electron, leads to statistical properties of the energy levels
of a new type, intermediate between the standard Poisson and GOE statistics.
Similar observations have been recently reported in other systems
\cite{bogo,rhombus,montambaux},
which indicates their importance and broad interest: as far
as we know, non-hydrogenic atoms are the first ``real" systems where
this is numerically observed and where an experimental confirmation
could be obtained in a near future.
 
We thank E. Bogomolny, K. Taylor and J. Zakrzewski for fruitful discussions.
CPU time has been provided by IDRIS.
Laboratoire Kastler-Brossel, de l'Ecole Normale Sup\'erieure et de
l'Universit\'e Pierre et Marie Curie, is unit\'e associ\'ee 18 du CNRS.

\end{document}